\documentclass[twocolumn,showpacs,preprintnumbers,amsmath,amssymb]{revtex4}

\usepackage{graphicx}% Include figure files
\usepackage{dcolumn}% Align table columns on decimal point
\usepackage{bm}% bold math

\begin{document}

\title{Comparative experimental study of local mixing of active and passive scalars in turbulent thermal convection}
\author{Quan Zhou and Ke-Qing Xia}
\address{Department of Physics, The Chinese University of Hong Kong,
Shatin, Hong Kong, China}
\date{\today}

\begin{abstract}

We investigate experimentally the statistical properties of active
and passive scalar fields in turbulent Rayleigh-B\'{e}nard
convection in water, at $Ra\sim10^{10}$. Both the local
concentration of fluorescence dye and the local temperature are
measured near the sidewall of a rectangular cell. It is found
that, although they are advected by the same turbulent flow, the
two scalars distribute differently. This difference is twofold,
i.e. both the quantities themselves and their small-scale
increments have different distributions. Our results show that
there is a certain buoyant scale based on time domain, i.e. the
Bolgiano time scale $t_B$, above which buoyancy effects are
significant. Above $t_B$, temperature is active and is found to be
more intermittent than concentration, which is passive. This
suggests that the active scalar possesses a higher level of
intermittency in turbulent thermal convection. It is further found
that the mixing of both scalar fields are isotropic for scales
larger than $t_B$ even though buoyancy acts on the fluid in the
vertical direction. Below $t_B$, temperature is passive and is
found to be more anisotropic than concentration. But this higher
degree of anisotropy is attributed to the higher diffusivity of
temperature over that of concentration. From the simultaneous
measurements of temperature and concentration, it is shown that
two scalars have similar autocorrelation functions and there is a
strong and positive correlation between them.
\end{abstract}


\maketitle

\section{Introduction}
\label{intro}

In the studies of hydrodynamic turbulence, scalar field has been
one of the main subjects besides the velocity field. This is
partly because a scalar field is presumed to be easier to tackle
with than a vector field, both experimentally and theoretically.
In general, there are two main kinds of scalar fields, classified
according to the interaction between the scalar and its carrier
flow: One is active scalar which couples dynamically to the
velocity field and the other is passive scalar whose feedback to
the flow is negligible. Although both are governed by the same
advection-diffusion equation, active and passive scalars belong to
different realms of mathematics and physics. Due to the dynamical
coupling with the advecting velocity field, the problem of active
scalars is a nonlinear one. On the other hand, passive scalars
obey a linear equation because of the absence of the feedback to
the advecting velocity field. This allows a fully theoretical
treatment of the problem, which has been carried out recently
\cite{ss00nature, falkovich01rmp}. To study the properties of both
scalar fields, the turbulent Rayleigh-B\'{e}nard convection (RBC),
a paradigm for studying buoyancy-driven turbulent flows
\cite{castaing89jfm, sigga94arfm, sreenivasan00nature, gl00jfm,
kadanoff01pt}, provides an ideal platform. This is because in this
system the temperature field itself is already an active scalar,
as it drives the turbulent flow via buoyancy, and, if a
fluorescent dye solution is injected into the flow, the
concentration field of the dye behaves as a passive scalar. The
object of the present experimental investigation is to address the
following questions: What are the relations, similarities, and
differences between the active and passive scalar fields advected
by the same turbulent flow?

%And (ii) what are the behaviors of the mixing and geometry of the
%passive scalar field in buoyancy-driven turbulence?

Passive scalars have been studied extensively for many years, and
the results have provided many insights into the turbulence
problem \cite{sreenivasan91prsla, ss00nature, warhaft00arfm,
falkovich01rmp}. However, there have been a limited number of
studies on the relations between the active and passive scalar
fields. Celani \emph{et al.} \cite{celani04njp} compared the
behaviors of active and passive scalars in four different
numerical systems \cite{celani02prla, celani02prlb, celani04njp}
by focusing on the issue of universality and scalings and their
results provide two possible scenarios: the first is that active
and passive fields in the same flow should share the same
statistics if the statistical correlations between scalar forcing
and carrier flow are sufficiently week; whereas in the second
scenario these two fields may behave very differently for systems
with strong correlations between the active scalar input and the
particle trajectories. These authors thus suggested that a
case-by-case study is needed. For thermal convection, Celani
\emph{et al.} \cite{celani02prla, celani04njp} showed that in 2D
numerical convective turbulence the two scalar fields have the
same even-order anomalous correlations, which are universal with
respect to the choice of external sources, and their intermittency
in such systems may be traced back to the existence of
statistically preserved structures. However, these authors also
argued that the equivalence of the statistics of active and
passive scalar fields depend crucially on the statistics of the
whole velocity field, which may not hold in 3D convective
turbulence. Ching \emph{et al.} \cite{ching03pre} also offered
numerical evidence in the context of simplified shell models for
turbulent convection to show that under generic conditions the
even-order correlation functions of active scalars can be
understood via the emerging theory of statistically preserved
structures of the passive scalar counterpart. For
magnetohydrodynamics, Celani \emph{et al.} \cite{celani02prlb,
celani04njp} showed that a passive scalar displays a direct
cascade towards the small scales while the active magnetic
potential builds up large-scale structures in an inverse cascade
process. Gilbert and Mitra \cite{gilbert04pre} further found that
active and passive fields have different scaling properties in the
framework of a shell model. Ching \emph{et al.} \cite{ching03pre}
argued that such different scaling behaviors are due to the fact
that the active equations possess additional conservation laws and
the zero modes of the passive problem is not the leading factor
that dominates the structure functions of the active field.
However, some of the above fundamental relations between active
and passive scalar fields have not been investigated
experimentally.

In an earlier study, Zhou and Xia \cite{zhou02prl} investigated
experimentally the mixing of an active scalar, i.e. the
temperature field, in thermal turbulence. Their results show that
the active scalar field has sharp fronts or gradients with similar
statistical properties as those found in passive scalars, such as
the saturation of structure function exponent and log-normal
distribution of the width of fronts. However, to the best of our
knowledge, there have been no experimental investigation to
directly compare the fundamental properties of active and passive
scalars that are driven by the same turbulent velocity field. In
this paper, we undertake such studies in a turbulent RBC system
and our results suggest that the active scalar possesses a higher
level of intermittency.

The remainder of this paper is organized as follows. In Sec.
\ref{exp}, we describe the convection cell used in the
experiments, details of the temperature measurement and the
technique of laser-induced fluorescence (LIF), and experimental
conditions and parameters. The experimental results are presented
and analyzed in Sec. \ref{res}, which is divided into five parts.
Sec. \ref{pass} and \ref{acti} discuss the statistical properties
of local concentration and temperature fluctuations, respectively.
In Sec. \ref{buoyantscale} we show that there is a certain buoyant
scale in turbulent thermal convection, above which buoyancy
effects are important and temperature is active. Sec.
\ref{smallscale} compares the levels of intermittency and isotropy
of the two scalars, both above and below the buoyant scale.  Sec.
\ref{corr_ctf} studies the cross-correlation between the two
scalars. We summary our findings and conclude in Sec. IV.

\section{Experimental setup and procedures}
\label{exp}

\subsection{Convection cell}
\label{rect_cell}

\begin{figure}
\center
\resizebox{1\columnwidth}{!}{%
  \includegraphics{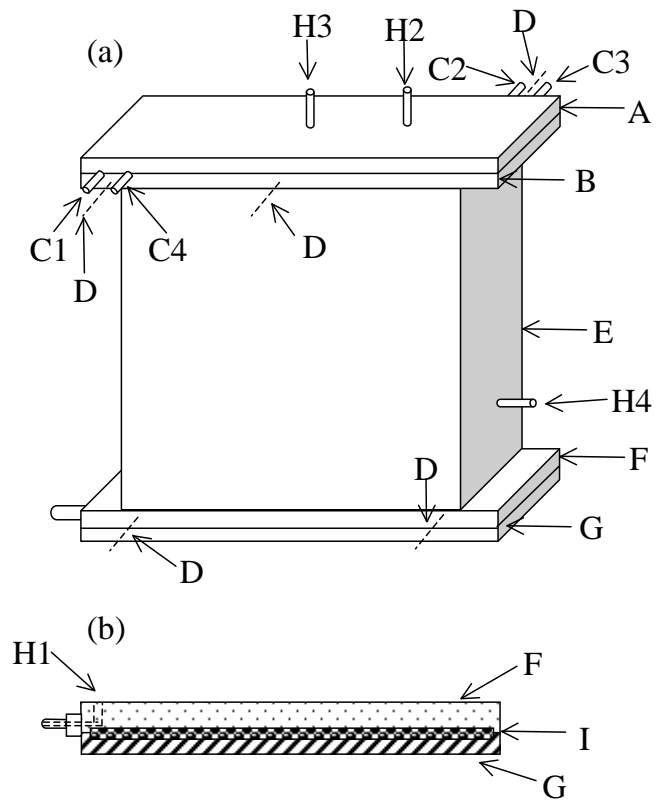}
} \caption{Schematic diagram of the rectangular cell. (a) Front
view of the rectangular cell and (b) cross-section view of the
bottom plate. A: the chamber cover, B: the top plate, C1, C2, C3,
C4: nozzles, D: thermistors, E: the Plexiglas sidewall, F: the
bottom plate, G: the cover for the bottom plate, H1: slit for dye
injection, H2: nozzle for dilution water injection, H3: nozzle for
inserting the small thermistor, H4: nozzle for fluid out-flowing,
I: heater.} \label{fig1}
\end{figure}

The scalar measurements were conducted in a rectangular turbulent
RBC cell, whose schematic drawing is shown in Fig. \ref{fig1}. The
length, width and height of the cell are $25.4\times7.5\times25.4$
cm$^3$, respectively, and thus the large-scale circulation (LSC)
is confined in the plane with the aspect ratio unity
\cite{xia03pre, zhou07pre, sun07jfm}. The sidewall of the cell,
indicated as E in the figure, is composed by four transparent
Plexiglas plates of thickness $7$ mm. The top (B) and bottom (F)
plates, whose surfaces are electroplated with Nickel to prevent
the oxidation by water, are made of copper with a thickness of
$1.5$ cm for its good thermal conductivity ($400$ W/mK). Rubber
O-rings are placed between the copper plates and the sidewall
plates to avoid fluid leakage. A water chamber is constructed with
the upper surface of the top plate and an attached stainless cover
(A). There are four nozzles (C1, C2, C3 and C4), located at the
end of the long edges, through which a refrigerated circulator is
connected to the chamber. To keep the temperature of the upper
plate uniform, water is pumped into the chamber through two
nozzles C1 and C3, circulates within the chamber, and then goes
out from the other two nozzles C2 and C4, respectively. A
rectangular film heater (I) is sandwiched between the lower copper
plate (F) and a stainless steel plate (G). There is a slit (H1),
through which the fluorescence dye solution is injected into the
cell, parallel to the short edge and $1$ cm from the sidewall on
the lower copper plate. To minimize the perturbation of the bottom
plate and the lower boundary layer, the area of the slit for
injection $S_{slit}$ is very small, which is $1$ mm in width and 2
cm in length, so the slit takes up only $0.1\%$ of the bottom
plate's total area. Diluted water is injected from the nozzle H2
to keep the background concentration of the cell unchanged and the
nozzle H4 is used for water out-flowing.

\subsection{Temperature measurements}
\label{temp_mea}

\begin{figure}
\center
\resizebox{1\columnwidth}{!}{%
  \includegraphics{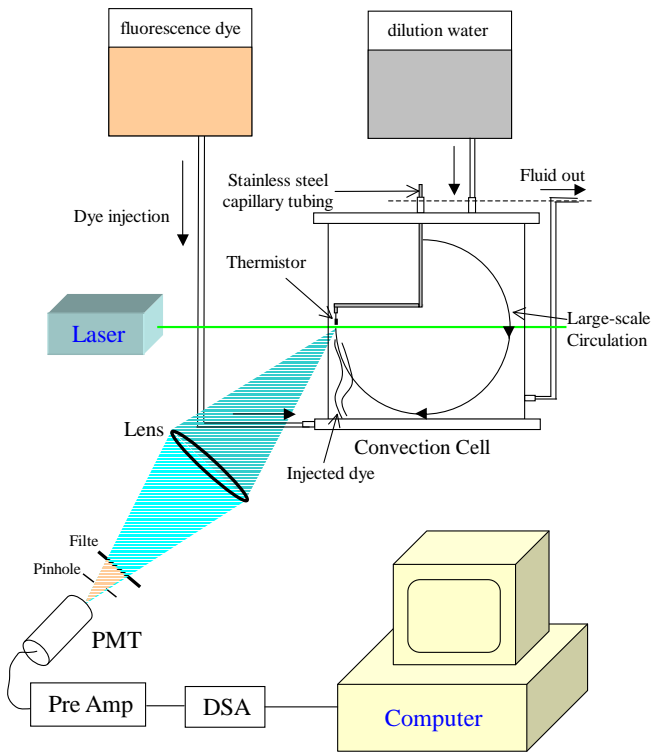}
}\caption{(Color online). Schematic diagram of the experimental
setup for simultaneous measurements of temperature and
concentration.} \label{fig2}
\end{figure}

There are two different types of thermistors used in our
experiments for the temperature measurements. Five thermistors of
the first type (Model 44031, Omega Engineering Inc.), indicated as
D in Fig. \ref{fig1}, are imbedded inside the two plates to
monitor their respective temperatures. Three thermistors are in
the top plate and the other two are in the bottom one. A
$6.5$-digit multimeter (Model 2000, Keithley Instruments Inc.) is
used to record the resistance of the thermistors with a sampling
frequency of $\sim1$ Hz. The measured relative temperature
difference between two thermistors in the same plate is found to
be less than $2\%$ of that across the convection cell for both
plates in our experimental conditions, indicating that the
temperature is uniform across the two horizontal plates. The
second type of thermistors (AB6E3-B10KA103J, Thermometrics) are
used in the local temperature measurement inside the convecting
fluid. They have a sensing head of $250$ $\mu m$ in diameter and a
thermal time constant of $10$ ms in water, which is fast enough to
detect the temperature fluctuations in thermal turbulence. As
shown in Fig. \ref{fig2}, the thermistor is threaded through a
stainless steel capillary tubing (U143, Fisher Scientific) with
outer diameter of $1.5$ mm and inner diameter of $1$ mm and then
attached to a translational stage with an accuracy to $0.001$ mm.
This arrangement allows the precise adjustment of the separation
between the temperature probe and the laser beam, which is used
for the local concentration measurement and will be introduced in
Sec. \ref{lif_mea}. A Wheatstone bridge with AC source and a
lock-in amplifier (Model SR830 DSP, Stanford Research Systems) are
used to measure the temperature fluctuations. A Dynamic Signal
Analyzer (DSA, HP 35670A) with a 24-bit dynamic range and four
input channels is used to digitize and record the signals and then
transmit them to computer.

\subsection{LIF technique}
\label{lif_tech}

There are many experimental techniques for measuring a passive
scalar field in turbulent flows, among which the LIF
(laser-induced fluorescence) technique seems to provide the best
non-intrusive method for the high-spatial-resolution measurement
of concentration. The use of this technique in fluid mechanics
have been well documented \cite{dimotakis85aiaa, walker87jpe,
sreenivasan90jfm, saylor95eif, dimotakis96jfm, crimaldi97eif,
wang00eif}. In almost of all these studies were carried out in
open systems, such as turbulent flows in a pipe or in a channel.
Whereas, LIF experiments involving longtime measurements in closed
systems, such as the turbulent RBC, have not been reported.
Longtime measurements in a close system will result in an increase
of mean concentration of the system and hence leads reductions in
the dynamical range of the fluorescence signal and also in the
signal-to-noise ratio. In this section we present the details of
the apparatus for our LIF measurements and how we overcome such
difficulties, and show the validation of this technique in our
case.

\subsubsection{Local concentration measurement}\label{lif_mea}

Figure \ref{fig2} shows the schematic diagram of the experimental
setup for local concentration measurement. The solution of the
fluorescence dye of Rhodamine 6G is injected from the bottom plate
of the convection cell, and then a laser light source is used to
illuminate the flow. From the absorption spectrum of dye solution
with concentration of $1.33\times10^{-7}$ M measured by a
monochromator (Model 300i, Acton Research Corporation) and an ICCD
(Model TE/CCD-1024-E/1, Princeton Instruments Inc.), one sees that
there is an absorbtion peak of fluorescence dye around the
wavelength of $520$ nm. Thus the green light with the wavelength
of $514.5$ nm from an argon ion laser (Coherent Innova 70) is
chosen as the exciting light. For low concentration, the resulting
fluorescence intensity at one point is proportional to the local
concentration. We use a photomultiplier tube (PMT, R2368,
Hamamatsu) to quantify the intensity of the fluorescence. From the
fluorescence spectrum of dye solution measured by the same
equipments as for the absorbtion spectrum, there is an emission
peak around the wavelength of 560 nm, so a filter with a passband
from 550 nm to 570 nm is chosen and is placed in front of the PMT
to block out the incident green light. To improve the
signal-to-noise ratio, the experiments were conducted in a dark
environment to minimize noise due to ambient light and a low-noise
preamplifier (Model SR560, Stanford Research Systems) is used to
filter out high frequency noise and amplify the signal obtained
from PMT. The four-channel DSA, which is also used for recording
the temperature fluctuations, digitizes and records the
fluorescence signals and then transmits them to computer.

Since the turbulent RBC is a closed system, a continuous injection
of dye will increase the background concentration level in the
convection cell and thus reduce the dynamical range of the
fluorescence signal. To solve this problem, we injected dilution
water into the cell to keep the background concentration constant.
As shown in Fig. \ref{fig2}, both the dye solution and dilution
water are supplied to the cell from the raised tanks and by the
pressure difference between the tanks and the cell and hence the
flow rate of injected fluids can be varied by changing the height
of the tanks. The out-flowing fluid is captured in a third tank
(not shown in the figure) from the nozzle H4 in Fig. 1. To
minimize the perturbation of the flow field in the convection
cell, the speed of injected fluids is approximately the same as
that of the large-scale circulation (LSC) of the convective flow,
which is about 1 cm/s in our case. In the present case, the volume
of injected fluids per unit time is very small (about $0.2$ mL/s
for the dye solution and about $0.4$ mL/s for the dilution water,
in contrast to the convecting fluid volume of $4.84$ L).
%
%The flow rate is calculated manually by collecting the out-flowing
%fluid at nozzle H4 over a minute period and determining its volume
%using a measuring cylinder. For the purpose of long-time
%measurements, the dye solution and dilution water are pumped into
%the first two tanks respectively every 30 minutes.

\subsubsection{Calibration for LIF}\label{lif_cali}

Previous works have pointed out that there are some effects which
can influence LIF measurement and hence change the measured
magnitude of the dye concentration \cite{dimotakis85aiaa,
walker87jpe, saylor95eif, crimaldi97eif, wang00eif}. The effects
which we must consider in our experiments are as follows:

\textbf{PH and temperature effects: }The measured fluorescence
intensity $I_f$ may vary with PH and temperature
\cite{walker87jpe}. However, in our experiment, the PH of the
fluid in the convection cell is almost constant ($\sim5.5$) and
the temperature fluctuations at the measuring position is about
$2^{\circ}C$. In this range, PH and temperature effects can be
neglected.

\textbf{Absorption along the illumination path: }For our optical
configuration, the incident laser must pass through the background
fluid which itself contains dye in it before illuminating the
measuring position. This causes the attenuation of the
illumination source along its path due to absorption and then a
non-linear relationship between fluorescence intensity and dye
concentration \cite{walker87jpe, crimaldi97eif}. However for
sufficiently low concentration, absorption effects along the
incident laser beam path can be neglected and the fluorescence
intensity at the measuring position can be expressed linearly as
\begin{equation}\label{eq:if}
I_f = \beta I_e C,
\end{equation}
where $\beta$ is a proportionality constant, $I_e$ is the
intensity of the induced laser and $C$ is the dye concentration.

\textbf{Photobleaching: }If the exciting laser intensity is high
enough or the velocity $v$ of the fluid is zero or too small for
the LIF measurement, some dye molecules, after excited, don't
absorb or re-emit light and the fluorescence intensity decays with
time, due to photodecomposition or collision quenching, etc. This
phenomenon is called photobleaching and have been well studied in
many experiments \cite{dimotakis85aiaa, saylor95eif,
crimaldi97eif, wang00eif}.

\begin{figure}
\center
\resizebox{1\columnwidth}{!}{%
  \includegraphics{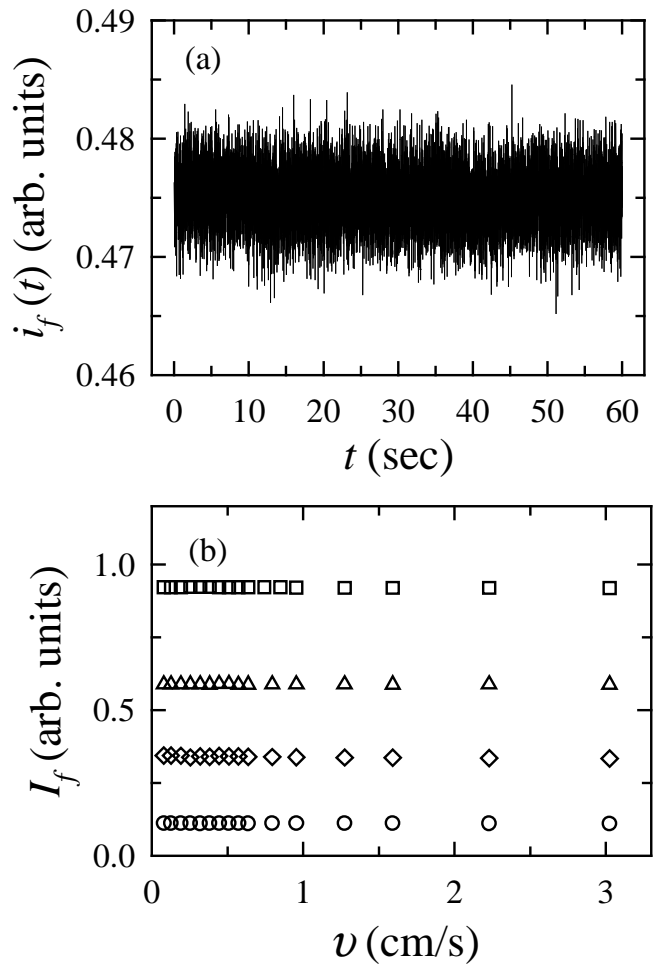}
}\caption{(a) A sample time series of $i_f(t)$ measured at the
flow cell center. Data taken at $C=2.0\times10^{-7}$ M, $I_e=200$
mW and $\upsilon=0.14$ cm/s. (b) $I_f$ ($=\langle i_f(t)\rangle$)
vs $\upsilon$ at $I_e=200$ mW for, from bottom to top,
$C=0.5\times10^{-7}$ M, $1.5\times10^{-7}$ M, $2.5\times10^{-7}$
M, and $4.0\times10^{-7}$ M.} \label{fig3}
\end{figure}

To determine the proper ranges in which our experiments can be
carried out properly, we calibrated the fluorescence intensity
$I_f$ under controlled conditions (e.g. $C$, $I_e$ and $v$). The
convection cell in Fig. \ref{fig2} was replaced by a flow cell and
a rectangular water tank, which contains dye solution with known
concentration, is used as a reservoir. Water was pumped from the
tank to the flow cell and then returning back to the tank by a
gear pump. The fluorescence intensity was measured at the center
point of the flow cell. Assuming a laminar velocity profile, we
can obtain the mean velocity at the measuring position from the
flow rate of the pump and the cross section area of the flow cell.

\begin{figure}
\center
\resizebox{1\columnwidth}{!}{%
  \includegraphics{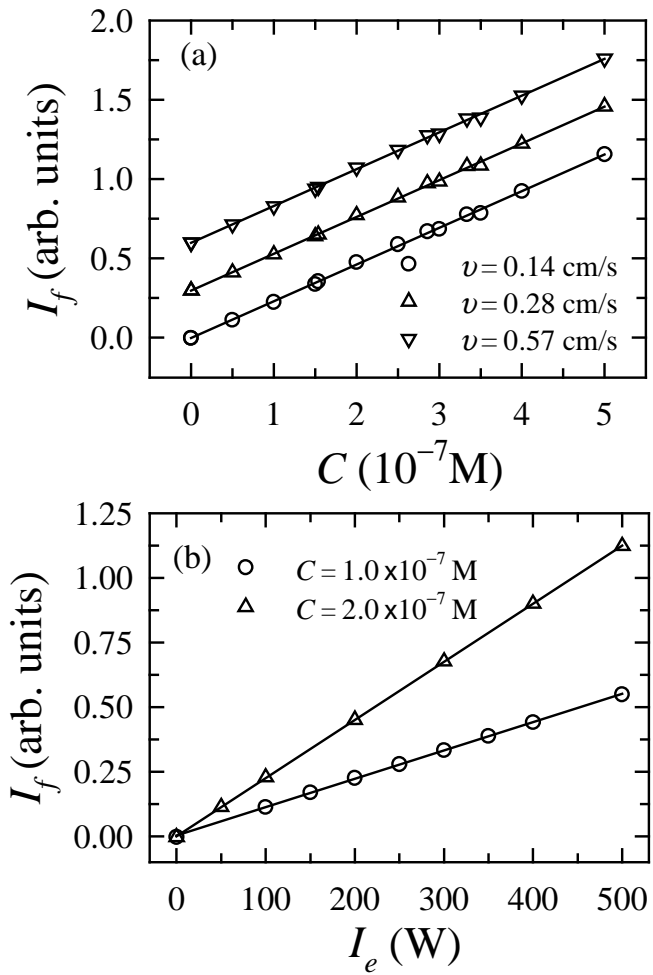}
}\caption{(a) $I_f$ vs $C$ at $I_e=200$ mW for three different
velocities. For clarity, the $\upsilon=0.28$ cm/s and $0.57$ cm/s
data have been shifted up by $0.3$ and $0.6$, respectively. (b)
$I_f$ vs $I_e$ at $\upsilon=0.28$ cm/s for two different
concentrations. The solid lines show the linear fits to the
corresponding data.} \label{fig4}
\end{figure}

A sample time-series of the measured instantaneous fluorescence
intensity $i_f(t)$ is shown in Fig. \ref{fig3}(a). One sees no
attenuation of $i_f(t)$ in this 60-second segment. In Fig.
\ref{fig3}(b), we show $I_f$ ($=\langle i_f(t)\rangle$, where
$\langle\cdots\rangle$ denotes a time average) measured at
different dye concentrations as functions of the flow velocity. We
see no appreciable dependence of $I_f$ on the fluid velocity $v$.
In our scalar experiment, the mean velocity at the measuring
position is about 1.8 cm/s at $Ra=1.2\times10^{10}$ (see Sec.
\ref{corr_ctf}). So in this range of fluid velocity photobleaching
effects can be neglected. In Fig. \ref{fig4}(a) we show measured
fluorescence intensity, at several flow velocities, vs the
concentration $C$. It is seen that, over the range measured
($<5.0\times10^{-7}$ M), $I_f$ is linear with $C$. In Fig.
\ref{fig4}(b) $I_f$ was measured with increasing incident laser
intensity $I_e$, which was measured by a power meter (model
1825-C, Newport). For $I_e<500$ mW, $I_f$ is found to be
proportional to $I_e$. Based on the above measurements, we chose
the concentration of the injected dye solution $C_0$ to be
4.0$\times10^{-7}$ M, the background concentration $C_{back}
=1.33\times10^{-7}$ M and the incident laser intensity $I_e=200$
mW in our scalar experiment. In these parameter ranges Eq.
(\ref{eq:if}) is valid and the fluorescence intensity $I_f$ is
proportional to the local concentration $C$. Thus the local
concentration can be obtained directly from the measured
fluorescence intensity.

\subsection{Experimental conditions and parameters}
\label{exp_con}

\begin{table*}
\centering \caption{Typical experimental parameters. $\bar{u}$ is
determined from the cross-correlation between temperature and
concentration fluctuations in Sec. \ref{corr_ctf}. $t_B$ and $t_0$
($\simeq 4H/\bar{u}$) are the Bolgiano time scale and the turnover
time of LSC, respectively.}
\begin{ruledtabular}
\begin{tabular}{c c c c c c c c}

$\Delta T$ & $H$ & Ra & Pr & Sc &  $Pe_T$ & $Pe_C$ & Nu \\
($^{\circ}$C) & (cm) & & & & & & \\
\hline

27.85 & 25.4 & $1.2\times10^{10}$ & 5.3   & 2100  &
$3.16\times10^4$ & $1.25\times10^7$ & 140\\ \hline \hline
$T_{bulk}$  & $\nu$ & $\kappa_T$ & $\kappa_C$ & $\bar{u}$ & $t_B$  & $t_0$ \\
($^{\circ}$C) & $(m^2/s)$ & $(m^2/s)$ & $(m^2/s)$ & (cm/s) & (sec) & (sec) &\\
\hline
30.97  & $7.85\times10^{-7}$ & $1.48\times10^{-7}$ & $3.73\times10^{-10}$ & 1.84 &  1.3  & 55.2 &\\

\end{tabular}
\end{ruledtabular}
\end{table*}

In our experiments, water is used as the convecting fluid and
Rhodamine 6G as the fluorescence dye. During the experiments, the
constant power is supplied to the bottom plate of the convection
cell, so that it is under a constant-flux boundary condition, but
at steady state its temperature remains effectively constant; the
top plate's temperature is regulated so that it is under
constant-temperature boundary condition. The control parameters
are the Rayleigh number Ra, the Prandtl number Pr and the Schmidt
number Sc, which are defined as,
\begin{equation}\label{eq:rapr}
Ra = \frac{\alpha g \Delta TH^3}{\nu \kappa} \ ,\mbox{\ }Pr =
\frac{\nu}{\kappa_T}\ , \mbox{\ and\ } Sc = \frac{\nu}{\kappa_C}\
,
\end{equation}
respectively. Here $\alpha$, $\nu$, $\kappa_C$, and $\kappa_T$ are
the coefficients of thermal expansion, kinematic viscosity,
molecular diffusivity of the fluorescence dye and thermal
diffusivity of the working fluid, respectively, $g$ is the
acceleration due to gravity, $\Delta T$ is the temperature
difference between the bottom and the top plates of the convection
cell, and $H$ is the height of the cell. Another parameter used to
characterize the rate of advection of the scalar by a flow to its
rate of diffusion is P\'{e}clet number, which is defined as,
\begin{equation}
Pe_T = \frac{\bar{u}H}{\kappa_T} \mbox{\ \ and\ \ } Pe_C =
\frac{\bar{u}H}{\kappa_C}
\end{equation}
for temperature and concentration, respectively. Here $\bar{u}$ is
the mean velocity at the measuring position. Table I summarizes
the typical experimental parameters of the present investigation.
From the table, one sees that, although $Pe_C$ is two orders
larger than $Pe_T$, both $Pe_C$ and $Pe_T$ are much larger than 1,
suggesting that the mixing of the temperature and concentration
fields are both dominated by the turbulent velocity field, rather
than their diffusion rates. Table I also shows that the
temperature difference across the cell $\Delta T$ is
27.85$^{\circ}$C. Previous studies \cite{ahlers05jfm, sun05jfm}
have pointed out that to strictly conform to the Boussinesq
condition (in water) $\Delta T$ should be limited to $\lesssim$ 15
$^{\circ}$C. Thus, there may exist some Non-Boussinesq effect in
our current system. Non-Boussinesq effect in water manifests as an
increase in the mean bulk temperature as compared to the average
of the top and bottom temperatures, and it also slightly reduces
the heat flux, i.e. Nusselt number \cite{ahlers06jfm}. However,
its effect on the temperature and concentration fluctuations are
unknown.

In addition to the injection from the bottom plate, the dye
solution was also injected into the central region of the
convection cell. Due to the absence of the mean flow in the cell
center, however, the dye packets meander around after injection
and we could not acquire sufficient amount of mixing data
(signals) for statical analysis. The measurements were also
carried out near the sidewall of a cylindrical cell. Comparable
features and behaviors of the passive scalar and of its relations
with the active scalar were obtained. However, due to azimuthal
meandering of the LSC in the cylindrical cell
\cite{sun05prl,ahlers05prl,xi06pre}, the injected dye solution
wobbles with the LSC and cannot always pass through the laser
beam, resulting in a poor signal-to-noise ratio. The statistics
are thus worse for the cylindrical cell. Therefore, all results
presented and discussed in this paper are from measurements made
near the sidewall of the rectangular cell with dye injected from
the bottom plate.

For the present Ra and Pr, the viscous boundary layer thickness
near the sidewall is about 2 mm \cite{qiu98pre}. Thus the
measuring position is chosen at $1$ cm from the sidewall and at
the mid-height of the convection cell, which is outside of the
sidewall viscous boundary layer and inside the LSC. The cell is
tilted with a small angle of about $1^{\circ}$ near the side of
the dye injection slit to lock the LSC and to make the dye
solution move upwards with the LSC after it is injected into the
cell.

To reveal the statistical properties of passive and active
scalars, longtime measurements of the local concentration and
temperature fluctuations were carried out independently. The
measurement of concentration lasted 27 hours with a sampling
frequency of 64 Hz. While the time series of temperature
fluctuations, which was obtained at the same measuring position
and under the same conditions as concentration, consists of three
parts, each of which lasted 20 hours and all were acquired at a
sampling rate of 64 Hz. To study the relations between these two
scalars, a simultaneous measurement of the local concentration and
temperature fluctuations was also made. In this case, the
temperature probe (thermistor) was placed above the concentration
probe (laser beam). The separation $\ell$ between two probes can
be adjusted by a translational stage. For each separation, we
recorded two hours of data, except the measurement for
$\ell=0.375$ mm which lasted 6 hours, with the sampling frequency
of 128 Hz for both concentration and temperature.

\section{Results and discussion}
\label{res}

\subsection{The passive scalar measurement}
\label{pass}

\begin{figure}
\center
\resizebox{1\columnwidth}{!}{%
  \includegraphics{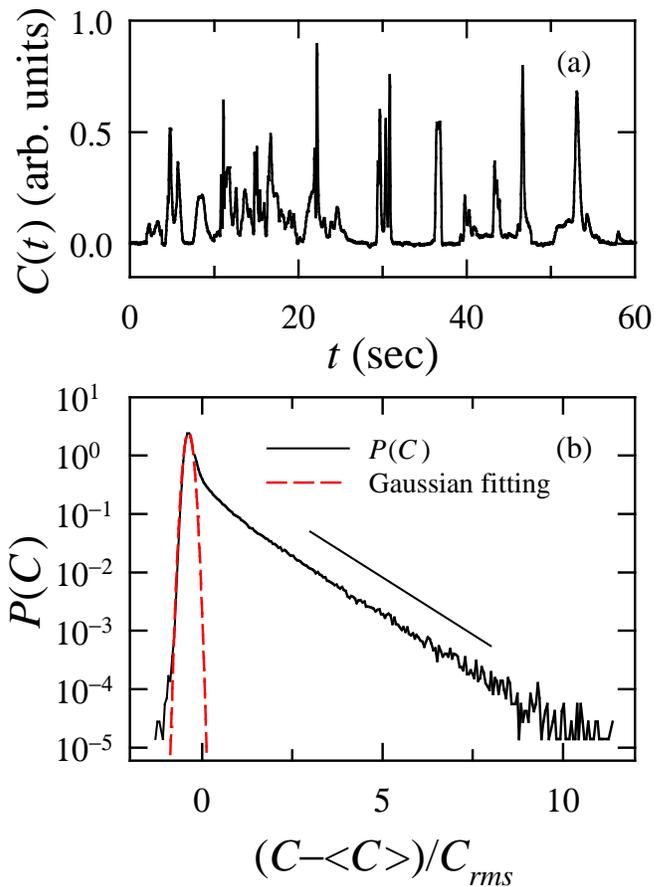}
}\caption{(Color online). (a) A 60-second time series of
concentration fluctuations. (b) PDF of concentration fluctuations
$P(C)$. The dashed curve represents a Gaussian-function fitting to
the background signal and the solid line is for reference.}
\label{fig5}
\end{figure}

Figure \ref{fig5}(a) shows a typical time series of the measured
concentration fluctuations. When the injected dye packets passing
through the concentration probe, a signal with a pulse-like shape
would be detected. From the figure, one sees that our measured
$C(t)$ consists of many such pulse-like signals of variable
heights, corresponding to dye packets with different concentration
levels.

In homogeneous turbulent flows, probability density function (PDF)
of the velocity fluctuations is Gaussian \cite{warhaft91prl} (It
was also showed that the velocity fluctuations in turbulent RB
system is Gaussian distributed \cite{qiu04pof, sun05pre}). It is
generally assumed that the PDF of a passive scalar is also
Gaussian, since it is advected by the Gaussian-distributed
turbulent velocity field. Earlier measurements about the scalar
PDF in homogeneous turbulence \cite{tavoularis81jfm} showed that a
Gaussian distribution was a satisfactory model for the scalar
signal itself except that the tails of the PDF have not been
examined. However, Jayesh $\&$ Warhaft \cite{warhaft91prl} found
that for a mean temperature gradient in grid turbulence the tails
of the temperature PDF have an exponential shape (in their system
temperature is passive). Our measured PDF $P(C)$ is shown in Fig.
\ref{fig5}(b). One sees that the PDF consists of two parts. The
left part, which has a abscissa value below $-0.25$, is the
distribution for the background signals. The distribution of this
part is approximately Gaussian, and one can use a Gaussian
function to fit the data of this part (the dashed curve in the
figure). The right part is the distribution of the mixed
fluorescence dye. It can be seen that this part exhibits an
exponential decay, which was also observed for passive scalars in
other types of turbulent flows, e.g. in a turbulent jet
\cite{villermaux98crass}.

\begin{figure}
\center
\resizebox{1\columnwidth}{!}{%
  \includegraphics{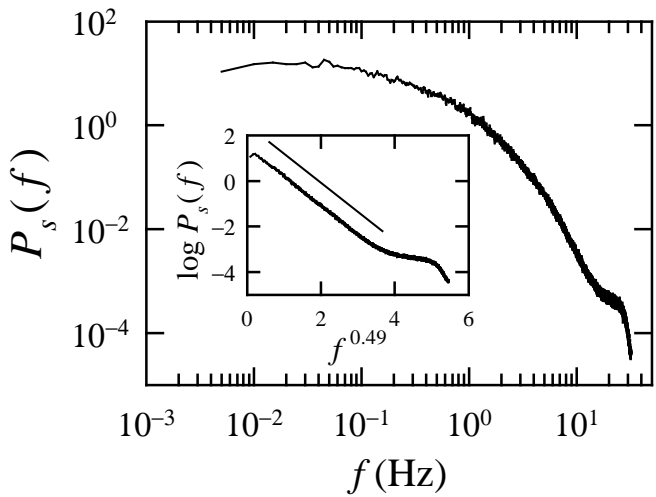}
}\caption{Power spectra $P_s(f)$ of the fluctuating concentration
$C(t)$. Inset is the plot of $\log[P_s(f)]$ as a function of
$f^{0.49}$ for the same data. The solid line is for reference.}
\label{fig6}
\end{figure}

Figure \ref{fig6} shows the power spectra $P_s(f)$ of the measured
dye concentration $C$. The figure shows that the magnitude of
$P_s(f)$ spans about five decades from the large scale to the
noise level. And we see no obvious scaling in the low-frequency
range, which may be due to a low Reynolds number (about 5700 in
our scalar experiments). At higher frequencies, the measured
$P_s(f)$ drops sharply and can be fitted by a stretched
exponential function
\begin{equation}
\label{ps_pass} P_s(f)\sim e^{\displaystyle{[-(f/f_c)^b}]},\mbox{\
\ } b=0.49\pm0.05
\end{equation}
where $f_c$ is a cutoff frequency. The inset of Fig. \ref{fig6}
shows a semi-logarithmic plot of the measured $P_s(f)$ as a
function of $f^{0.49}$. In this plot, the stretched exponential
function appears as a straight line. The cut-off frequency of the
concentration is around $10\sim20Hz$, which approximately equals
to that of the temperature (see Sec. \ref{acti}).

\subsection{The active scalar measurement}
\label{acti}

\begin{figure}
\center
\resizebox{1\columnwidth}{!}{%
  \includegraphics{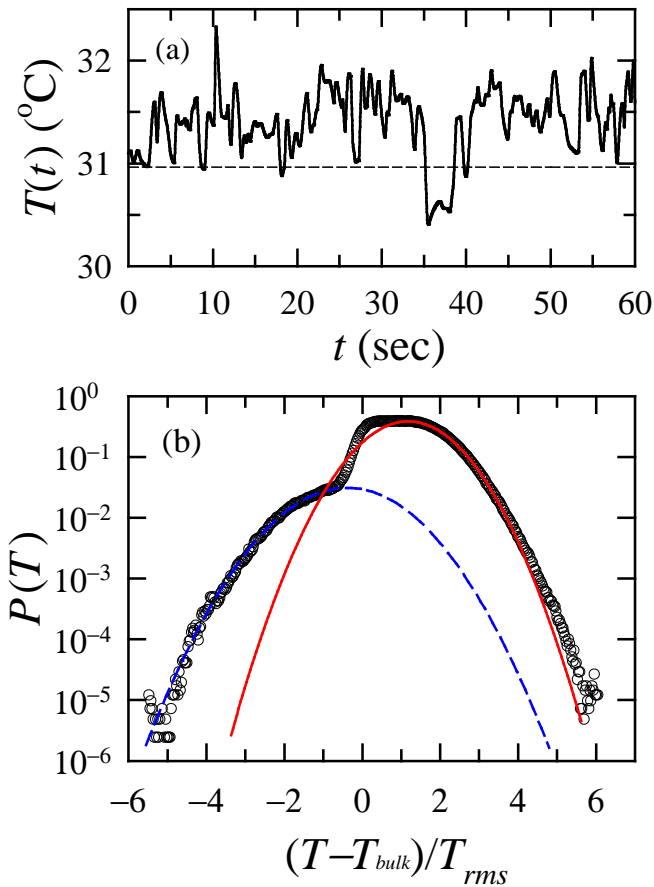}
}\caption{(Color online). (a) A 60-second time series of
temperature fluctuations. The dashed line indicates the mean bulk
temperature $T_{bulk}$. (b) Corresponding PDF of normalized
temperature fluctuations $P(T)$. The dashed curve and the solid
curve are two Gaussian-function fittings to the left and right
tails, respectively.} \label{fig7}
\end{figure}

Figure \ref{fig7}(a) shows how the temperature fluctuations change
with time. The measured $T(t)$ shows many intermittent upward
sharp spikes of variable heights, which may be attributed to the
upward-moving hot plumes. The corresponding PDF of the normalized
temperature fluctuations is shown in Fig. \ref{fig7}(b). Unlike
the exponential distributions observed in the center and near the
sidewall of a cylindrical cell, the measured PDF near the sidewall
in the rectangular cell shows a flat central region and two
Gaussian-like tails, which are fitted by separate Gaussian
functions (the solid and dashed curves). The Gaussian-like
distribution of temperature fluctuations means that the
temperature probe felt a lot of thermal coherent structures, i.e.
thermal plumes, passing through it \cite{gollub91pra}. The ratio
of the long side to the short side of our rectangular convection
cell is about $3.3$, so that the LSC is largely locked in the
vertical plane parallel to the long side. Hence most of thermal
plumes would group together, move upwards with the LSC and pass
through the measuring position due to the confinement in the other
direction. Thus, although our experiments with
$Ra=1.2\times10^{10}$ were carried out in the so-call ``hard
turbulence" regime \cite{castaing89jfm}, the measured temperature
fluctuations near the sidewall have two Gaussian-like tails rather
than exponential ones. It is also not surprising that the right
peak is more than ten times higher than the left one, since the
LSC has been locked and at the measuring position there are more
hot plumes moving upwards than cold one moving downwards. Note
also that there is only one group of cold plumes in the 60-second
time series shown in Fig. \ref{fig7}(a), as compared to the large
number of hot plumes in the same plot.

\begin{figure}
\center
\resizebox{1\columnwidth}{!}{%
  \includegraphics{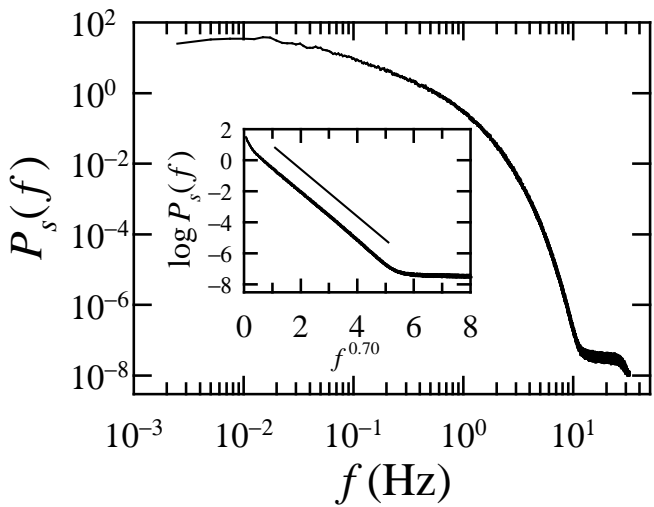}
}\caption{Power spectra $P_s(f)$ of the fluctuating temperature
$T(t)$. Inset is the plot of $\log[P_s(f)]$ as a function of
$f^{0.70}$ for the same data. The solid line is for reference.}
\label{fig8}
\end{figure}

We now turn to the power spectra of temperature fluctuations.
Based on dimensional arguments, it is traditionally thought that
there is simple scaling relation between $P_s(k)$ and $k$ in the
low-frequency region, i.e.,
\begin{equation}
P_s(k)\sim k^{-a},
\end{equation}
where $k$ is the wavenumber. Here, $a$ would be $5/3$ according to
Kolmogorov's arguments (K41)\cite{kolmogorov1941a,
kolmogorov1941b, O49, C51}, and it would be $7/5$ if the arguments
of Bolgiano \cite{B59} and Obukhov \cite{O59} (BO59) are correct.
However, from a direct multipoint measurement of the temperatrue
field, Sun \emph{et al.} \cite{sun06prl} showed that $a=8/5$ near
the sidewall of a cylindrical cell, which could be explained by a
simple scaling analysis based on the coaction of buoyancy and
inertial forces.

Figure \ref{fig8} shows the measured temperature power spectra
$P_s(f)$ as a function of $f$. From the figure, one sees that the
magnitude of $P_s(f)$ spans more than nine decades from the large
scale to the noise level. However, the scaling range in the
low-frequency region is so limited that one cannot tell the
accurate value of $a$. This may be due to the asymmetry of the
rectangular cell and the sidewall effect. Another possible reason
for this limited scaling range is that the measured $P_s(f)$ here
is based on time or frequency domain. By using the Taylor
frozen-turbulence hypothesis with the condition that the turbulent
velocity fluctuations is much smaller than the mean flow velocity,
one can relate the time domain results to those of the space
domain. However, it is known that the mean flow velocity is
comparable to the rms velocity near the sidewall in turbulent RBC
convection cell \cite{shang01pre}. This would flatten the measured
$P_s(f)$ in the low-frequency region. At higher frequencies, the
measured $P_s(f)$ drops sharply and can be fitted by a stretched
exponential function
\begin{equation}
P_s(f)\sim e^{\displaystyle{[-(f/f_c)^b}]},\mbox{\ \ }
b=0.70\pm0.05
\end{equation}
where $f_c$ is a cutoff frequency. (We also used the data measured
in a $81\times20\times81$-cm$^3$ rectangular cell from Ref.
\cite{zhou07pre} to study the temperature power spectra near the
sidewall and $b=0.73\pm0.05$ was obtained.) Previous measurements
of temperature in smooth \cite{wu90prl} and rough \cite{tong01pre}
cylindrical cells also showed a stretched exponential decay at
higher frequencies. However, the values of $b$ in those studies
are both $0.55\pm0.05$, which are smaller than our results. Power
spectrum of the temperature fluctuations describe the cascade of
entropy of the system and hence a larger value of $b$ here means a
more rapid cascade of the system's entropy compared with the
cylindrical convection cell. This may be due to the asymmetry of
the rectangular cell and the sidewall effect. The inset of Fig.
\ref{fig8} shows a semi-logarithmic plot of the measured $P_s(f)$
as a function of $f^{0.70}$. In this plot, the stretched
exponential function appears as a straight line. From Fig.
\ref{fig8}, one sees that the cut-off frequency of the temperature
is around $10\sim15Hz$, which is about the same as that of
concentration as discussed in Sec. \ref{pass}.

\subsection{The buoyant scale of convective turbulence}\label{buoyantscale}

In 1959, Bolgiano \cite{B59} and Obukhov \cite{O59} proposed that
there is a characteristic length scale in buoyancy-driven
turbulence, now commonly referred to as the Bolgiano scale
$\ell_B$, above which buoyancy effects are important. It has been
shown \cite{my75} that the Bolgiano scale can be estimated as
\begin{equation}
\label{eq:lb}
\ell_B=\frac{\epsilon_v^{5/4}}{\epsilon_T^{3/4}(g\alpha)^{3/2}},
\end{equation}
where $\epsilon_v$ and $\epsilon_T$ are the kinematic and thermal
dissipation rates. Note that if the local dissipation rates
$\epsilon_v(\vec{r})$ and $\epsilon_T(\vec{r})$ are used in the
above equation, then it gives a local Bolgiano scale. The global
Bolgiano scale may be obtained by using the expressions of the
globally averaged dissipation rates $\epsilon_v$ and $\epsilon_T$
in terms of Ra, Pr, and Nu \cite{sigga94arfm}, which gives:
\begin{equation}
\label{eq:lb2} \ell_B=\frac{Nu^{1/2}H}{(RaPr)^{1/4}},
\end{equation}
As all our measurements are made in the time domain, we use $t_B$,
which is the counterpart of $\ell_B$ in the time domain, as a
characteristic time scale in our system, i.e.
\begin{equation}
\label{eq:tb} t_B=\frac{Nu^{1/2}t_0}{(RaPr)^{1/4}},
\end{equation}
which can be easily evaluated from the measured values of Ra, Pr,
Nu and the large-scale flow turnover time $t_0$.

\begin{figure}
\center
\resizebox{1\columnwidth}{!}{%
  \includegraphics{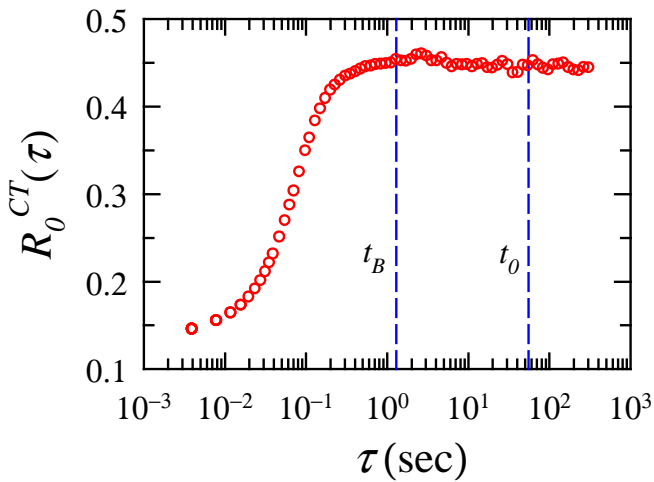}
}\caption{(Color online). Cross-correlation coefficient
$R_0^{CT}(\tau)$ of the concentration and temperature increments.
The two dashed lines show $t_B$ and $t_0$. The separation $\ell$
between  the temperature probe and the laser beam is $0.375$ mm.}
\label{fig9}
\end{figure}

To check whether this buoyant scale exists in turbulent RBC, we
compute the cross-correlation coefficient $R_0^{CT}(\tau)$ between
the increments of concentration and temperature, i.e.,
\begin{equation}
\label{eq:rct}
R_0^{CT}(\tau)=\frac{\langle\delta_{\tau}C(t)\delta_{\tau}T(t+\ell/\bar{u})\rangle}{\delta_{\tau}C(t)_{rms}\delta_{\tau}T(t)_{rms}}
\end{equation}
where $\delta_{\tau}C(t)=C(t+\tau)-C(t)$,
$\delta_{\tau}T(t)=T(t+\tau)-T(t)$, and $\ell/\bar{u}$ describes
the time delay between the concentration and temperature signals
due to the separation of the two probes. When the concentration
and temperature increments are uncorrelated, we have
$R_0^{CT}(\tau)=0$, while $R_0^{CT}(\tau)=1$ is obtained for a
linear relation between the two increments.

Figure \ref{fig9} shows the measured $R_0^{CT}(\tau)$ as a
function of $\tau$ for $\ell=0.375$ mm, which is the measured
separation between the laser beam and the thermistor tip. Two
dashed lines in the figure show $t_B$ and $t_0$, which were
determined from the measured Ra, Pr, and Nu (see Table I) using
Eq. (9) and from $4H/\bar{u}$ respectively. One sees that
$R_0^{CT}(\tau)$ increases with increasing $\tau$ when $\tau$ is
small and saturates at a value of $\sim0.45$ when $\tau\gtrsim
t_B$. It has been shown previously \cite{ching04jot} that the
vertical velocity increments correlate strongly with the
temperature increments above the Bolgiano time scale $t_B$,
reflecting strong effects of buoyancy. Since the concentration
fluctuations are dominated by the vertical velocity field near the
sidewall, a large positive correlation between the concentration
and temperature increments is expected and indeed observed here.
This supports the idea that, above the Bolgiano time scale,
buoyancy effects are significant and temperature can be regarded
as an active scalar. In contrast, for time scales below $t_B$,
buoyancy is not important and the temperature field behaves as a
passive scalar. Thus, the concentration and temperature
fluctuations are two separate stochastic fields at scales below
$t_B$. In this case, the correlation between them should be very
weak (as shown in Fig. \ref{fig9}, $R_0^{CT}(\tau)$ decreases
sharply when $\tau<t_B$), even though they are advected by the
same velocity field.

The physical picture revealed by Fig. \ref{fig9} is that a certain
buoyant scale exists in turbulent RBC, above which buoyancy
effects are important. Thus, the temperature field can indeed be
regarded as an active scalar above $t_B$, and as a passive scalar
below $t_B$.

\subsection{Statistical properties of scalar fluctuations}\label{smallscale}

\begin{figure}
\center
\resizebox{1\columnwidth}{!}{%
  \includegraphics{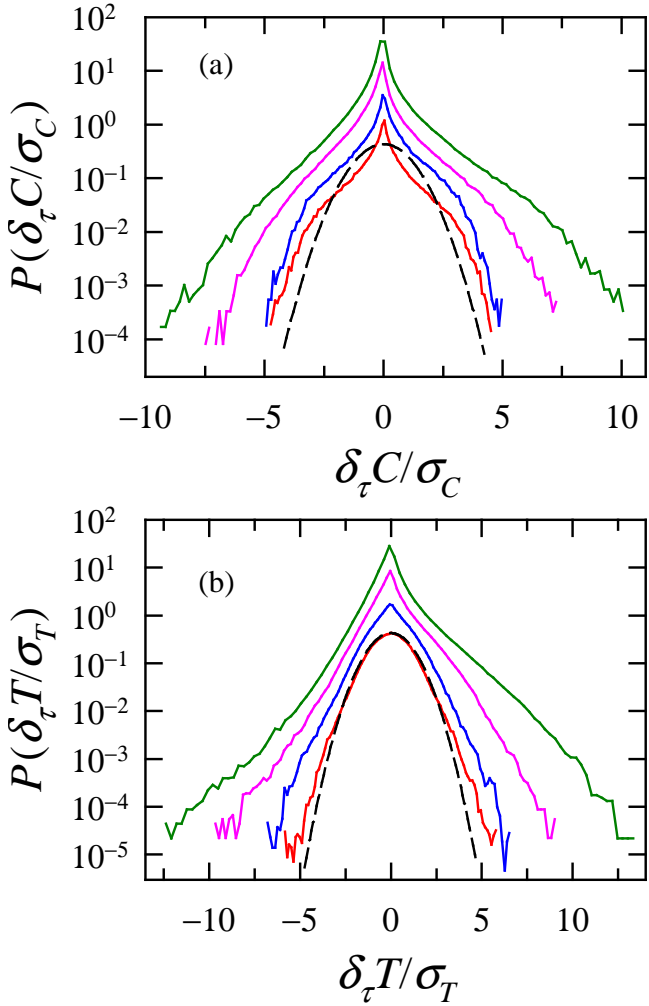}
}\caption{(Color online). PDFs of (a) concentration and (b)
temperature increments normalized by their standard deviations for
four time intervals $\tau/t_B = 0.05$, 0.15, 1, and 10, from top
to bottom. The curves have been shifted vertically for clarity and
the black dashed curve indicates a Gaussian distribution of
variance 1.} \label{fig10}
\end{figure}

We have seen in Sec. \ref{pass} that the right tail of
concentration PDF, which describes the distribution of the mixed
fluorescence dye, has a decreasing exponential shape. An
exponential distribution near the tail means higher probability of
rare events, i.e. intermittency, in comparison with the Gaussian
distribution. The small-scale intermittency of concentration
fluctuations can be characterized by the distributions of their
increments over different scales. Figure \ref{fig10}(a) shows the
measured PDFs of the concentration increments $\delta_{\tau}C(t)$
for four different time lags $\tau$, with $\tau$ = $0.05t_B$
(close to the dissipative scale), $0.15t_B$, $t_B$ (Bolgiano
scale), and 10$t_B$ ($\sim t_0/4$), from top to bottom. The PDFs
have been normalized by their own standard deviations
$\sigma_C=\langle (\delta_{\tau}C)^2\rangle^{1/2}$ and shifted
vertically for clarity. The dashed line in the figure represents a
Gaussian distribution of variance 1 for reference. Three features
are worthy of note: (i) all PDFs are strongly non-Gaussian from
the smallest time interval to lags of the order of the large
scale; (ii) all PDFs are clearly not self-similar, indicating
strong intermittency; (iii) there is a slight asymmetry of the
PDFs' tails, especially for small time lags, reflecting the
well-known property of small-scale persistence of anisotropy, i.e.
ramp-cliff structures \cite{sreenivasan91prsla, ss00nature,
warhaft00arfm}.

The distributions of the normalized temperature increments
$\delta_{\tau}T/\sigma_T$ ($\sigma_T=\langle (\delta_{\tau}T)^2
\rangle^{1/2}$) for the same values of $\tau$ as those for the
concentration are also examined, and are plotted in Fig.
\ref{fig10}(b). The small-scale persistence of anisotropy can also
be found for the active scalar increments from their asymmetric
distributions at small time lags. This asymmetry can be linked to
the sharp fronts of coherent structures in turbulent thermal
convection, i.e. thermal plumes \cite{zhou02prl}. Another notable
feature is that as time lags increase towards the large time scale
the shapes of the PDFs evolve from approximately exponential tails
to nearly Gaussian-like distribution, reflecting strong
intermittency as well.

When comparing the PDFs of concentration and temperature
increments, two features are remarkable: (i) The PDFs of
temperature increments at small scales [the top two curves in Fig.
\ref{fig10}(b)] are more asymmetric than those of concentration
increments at the corresponding scales, suggesting that the mixing
of temperature are more anisotropic than those of concentration.
(ii) Qualitatively, Fig. \ref{fig10} shows that the PDFs of
temperature increments change from non-Gaussian in small scales to
near Gaussian in large scales, whereas, those of concentration
increments remain non-Gaussian for all scales investigated. In
this sense, it suggests that the shapes of the concentration
increments' PDFs are less dissimilar among themselves than those
of the temperature increments' PDFs as time lags $\tau$ increase
towards the large time scale, and hence the passive scalar is less
intermittent than the active one \cite{pinton07prl}, which can
also be quantified by the flatness discussed below.

\begin{figure}
\center
\resizebox{1\columnwidth}{!}{%
  \includegraphics{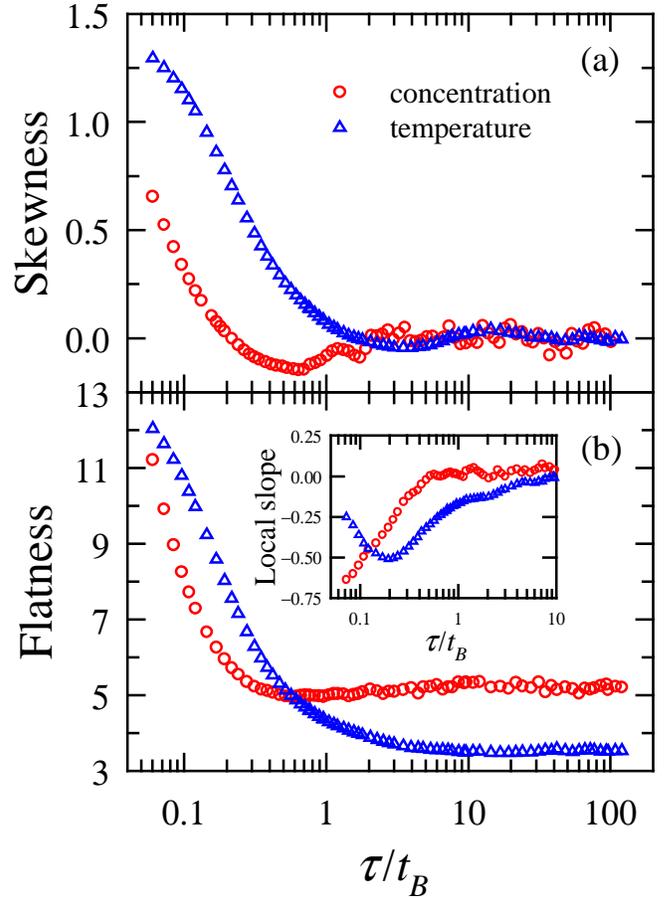}
}\caption{(Color online). (a) Skewness and (b) flatness of the
concentration (red circles) and temperature (blue triangles)
increments as functions of $\tau/t_B$. Inset of (b): the local
slopes, $\chi^X(\tau/t_B)=d \log{(F_X(\tau/t_B))}/ d
\log{(\tau/t_B)}$, as functions of $\tau/t_B$.} \label{fig11}
\end{figure}

To quantitatively characterize the anisotropic effects of the
small-scale mixing of the two scalars, we study the skewness of
their increments, i.e.,
\begin{equation}
S_X(\tau)=\langle(\delta_\tau X)^3\rangle/\langle(\delta_\tau
X)^2\rangle^{3/2}.
\end{equation}
where $\delta_{\tau}X(t)=X(t+\tau)-X(t)$ with $X=C$ for
concentration and $X=T$ for temperature. The skewness is a global
measure of asymmetry. By definition, the skewness of a
symmetric-distributed quantity is 0. Figure \ref{fig11}(a) shows
the measured skewness for the concentration and temperature
increments. One sees that both $S_C(\tau)$ and $S_T(\tau)$
decrease with time lags for small $\tau$. $S_C(\tau)$ reaches its
minimum value around $\tau\simeq0.7t_B$ and then increases a
little, whereas, $S_T(\tau)$ decreases almost monotonically to the
zero value. For $\tau\lesssim t_B$, both temperature and
concentration behave as passive scalars but one finds
$|S_T(\tau)|>|S_C(\tau)|$, suggesting that the mixing of
temperature are more anisotropic than those of concentration. This
is because the mixing of a passive scalar field becomes more
isotropic with decreasing diffusivity \cite{sreenivasan03prl} and
from Table I we see that the thermal diffusivity $\kappa_T$ is
about two orders of magnitude larger than molecular diffusivity
$\kappa_C$ of the dye. For $\tau\gtrsim t_B$, both $S_C(\tau)$ and
$S_T(\tau)$ fluctuate around 0, implying that the mixing of both
scalar are isotropic above the buoyant scale. This is strange
since buoyancy effects act on fluids along the vertical direction
but do not induce anisotropy for both temperature and
concentration.

We now examine more quantitatively the intermittency levels of the
two scalar fields. In general, the deviations of scaling exponents
of structure functions, $R^X_p(\tau)=\langle
|\delta_{\tau}X|^p\rangle\sim\tau^{\xi^X_p}$, in the inertial
range from those expected by the arguments of Kolmogorov
\cite{kolmogorov1941a, kolmogorov1941b} or Bolgiano \cite{B59} and
Obukhov \cite{O59} can be used as a measure for the degree of
intermittency. Unfortunately, above the Bolgiano time scale $t_B$,
no discernible scaling for concentration is observed no matter
whether structure functions are plotted directly against time lags
in a log-log plot or indirectly using an extended self-similarity
method \cite{benzi93a, benzi93b}. Here, we use the flatness of the
temperature and concentration increments, defined as,
\begin{equation}
F_X(\tau)=\langle(\delta_\tau X)^4\rangle/\langle(\delta_\tau
X)^2\rangle^2=R^X_4/(R^X_2)^2\sim\tau^{\chi^X},
\end{equation}
to characterize their levels of intermittency, where
$\chi^X=\xi^X_4-2\xi^X_2$. The flatness is a characteristic
measure of whether the data are peaked or flat relative to a
Gaussian distribution and hence can be used as a relatively
sensitive measure of intermittency. By definition, the flatness of
a Gaussian-distributed quantity is 3. In addition, the local slope
of the flatness for the temperature and concentration increments,
defined as
\begin{equation}
\chi^X(\tau) = d \log{(F_X(\tau)) / d \log{\tau}},
\end{equation}
may also be used as a measure of self-similarity, since
$\chi^X(\tau)=0$ within a certain range of scales for a quantity
whose increments are scale-free within that range. Therefore,
larger deviations of $\chi^X(\tau)$ from 0 reflects less
self-similarity, i.e. more intermittent. Figure \ref{fig11}(b)
shows the measured $F_C(\tau)$ and $F_T(\tau)$ plotted as
functions of time lags $\tau$. One sees that both quantities
decrease with increasing $\tau$ for small $\tau$ and cross at
$\tau\simeq0.5t_B$. For $\tau>10t_B$, $F_T(\tau)$ levels off to a
value of about 3.5, which is very close to that of a
Gaussian-distributed quantity, while $F_C(\tau)$ levels off at a
value of 5 for $\tau>0.5t_B$. The local slopes of $F_C(\tau)$ and
$F_T(\tau)$ are plotted as functions of $\tau$ in the inset of
Fig. \ref{fig11}(b). Above $t_B$, although one finds that the
value of $F_T(\tau)$ is smaller than that of $F_C(\tau)$,
temperature possesses a higher level of intermittency. This is
because $\chi^C(\tau)\simeq0$, suggesting an approximate
self-similar property, and $|\chi^T(\tau)|>|\chi^C(\tau)|$, i.e.
the deviations of $\chi^T(\tau)$ from 0 are larger than those of
$\chi^C(\tau)$. Recall that within such regime buoyancy is
important and the temperature field is an active scalar.
Therefore, our results presented here suggest that the active
scalar has a higher level of intermittency than the passive
scalar, at least in turbulent thermal convection. Below $t_B$, the
situation is more complicated. For $0.15t_B\lesssim\tau\lesssim
t_B$, $|\chi^T(\tau)|>|\chi^C(\tau)|$ is obtained, again
suggesting that temperature possesses a higher level of
intermittency within such regime. Whereas, for
$\tau\lesssim0.15t_B$, we have $|\chi^T(\tau)|<|\chi^C(\tau)|$,
suggesting that concentration is more intermittent for such
regime. Note that this regime is close to the dissipative scale,
but we don't understand the mechanism that caused such result.

\subsection{Cross-correlation between the concentration and temperature
fluctuations}\label{corr_ctf}

\begin{figure*}
\begin{minipage}[c]{.7\textwidth}
\centering
\includegraphics[scale=0.76]{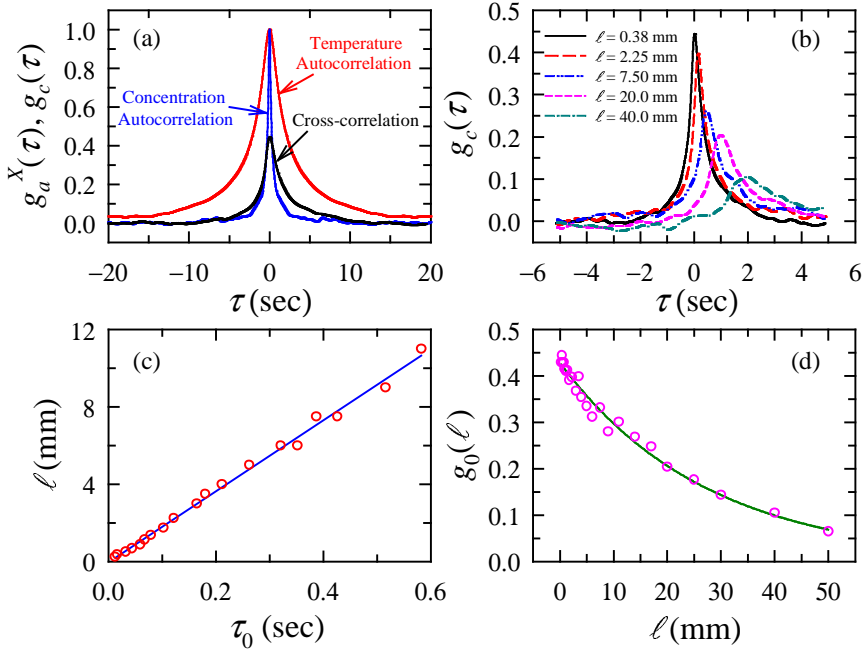}
\end{minipage}
\begin{minipage}[c]{.25\textwidth}
\centering
 \caption{(Color online). (a) Autocorrelation functions $g_a^T(\tau)$
(red line), $g_a^C(\tau)$ (blue line) and cross-correlation
function $g_c(\tau)$ (black line). The separation between the
laser beam and the temperature probe is $\ell=0.375$ mm. (b)
Cross-correlation functions $g_c(\tau)$ for five different
separations $\ell$. (c) The separation $\ell$ vs the position of
the measured correlation peak $\tau_0$. The solid line shows the
fitting linear function $\ell=\bar{u}\tau_0$ with $\bar{u}=18.4$
mm/s. (d) The measured correlation amplitude $g_0(\ell)$ as a
function of the separation $\ell$. The solid curve shows the
fitting exponential function
$g_0(\ell)=0.429e^{\displaystyle{-\ell/\xi}}$, with $\xi=27.3$
mm.} \label{fig12}
\end{minipage}
\end{figure*}

Figure \ref{fig12}(a) shows the measured auto-correlation
functions,
\begin{equation}\label{eq:auto_T}
g_a^X(\tau) = \frac{\displaystyle{\langle\delta X(t)\delta
X(t+\tau)\rangle}}{\displaystyle{X_{rms}^2}},
\end{equation}
for temperature ($X=T$) and concentration ($X=C$), respectively,
and their cross-correlation function
\begin{equation}\label{eq:cross}
g_c(\tau) = \frac{\displaystyle{\langle\delta C(t)\delta
T(t+\tau)\rangle}}{\displaystyle{C_{rms}T_{rms}}}
\end{equation}
as functions of time lag $\tau$, where $\delta X(t) = X(t) -
\langle X(t)\rangle$. One sees that all three functions have a
shape similar to each other, but the full width at half maximum
(FWHM) of their peaks are different. The FWHMs of the temperature
and concentration fluctuations are $W_T\simeq3.7s$ and
$W_C\simeq0.5s$, respectively, and the FWHM of their
cross-correlation is $W_{TC}\simeq2.2s$, which is about the
average of $W_T$ and $W_{C}$. One half of the FWHM of the
correlation functions, $W_T/2$ ($W_{C}/2$), provides information
about the transient time for the temperature (concentration)
fluctuations of a certain size passing through the temperature
(concentration) probe with a certain speed, and thus one can
obtain the coherent length of the temperature (concentration)
fluctuations from the FWHM and the advected speed. Since both
scalars are transported by the same velocity field and the
relation $W_T>W_C$ is obtained, we conclude that the coherent
length of the active scalar, i.e. the size of thermal plumes, is
larger than that of the passive scalar, i.e. the width of the
cliff structures of the concentration field
\cite{sreenivasan91prsla, ss00nature, warhaft00arfm,
tabeling01prl}, which is probably due to the larger value of the
thermal diffusivity.

Figure \ref{fig12}(b) shows the measured concentration-temperature
cross-correlation functions for five different sparation $\ell$.
It is found that the measured $g(\tau)$ all have a single peak
$g_0$ as concentration and temperature are both advected by the
same velocity field. However, the peak position $\tau_0$ for
different separation $\ell$ are not the same. The bigger $\ell$
is, the larger $\tau_0$ becomes. How the measured $\tau_0$ varies
with the separation $\ell$ is plotted in Fig. \ref{fig12}(c). It
can be clearly seen that $\ell$ increases with increasing
$\tau_0$, and the increasing manner may be described by a simple
linear function
\begin{equation}
\ell=\bar{u}\tau_0,\mbox{\ \ } \bar{u}=18.4 \mbox{\ mm/s}
\label{eq:linear_f}
\end{equation}
The solid line in Fig. \ref{fig12}(c) shows the fitting linear
equation (\ref{eq:linear_f}). The slope $\bar{u}=18.4$ mm/s,
describing the mean velocity with which the temperature and
concentration fluctuations pass through the measuring points. Fig.
\ref{fig12}(d) shows the relationship between the obtained $g_0$
and the separation $\ell$. It is clearly shown that $g_0(\ell)$
decreases with increasing $\ell$, and the decay may be described
by a simple exponential function
\begin{equation}
g_0(\ell)=0.429e^{\displaystyle{-\ell/\xi}},\mbox{\ \ } \xi=27.3
\mbox{\ mm} \label{eq:exp_f}
\end{equation}
The solid curve in Fig. \ref{fig12}(d) shows the fitting equation
(\ref{eq:exp_f}). A similar correlation length was also obtained
in velocity-temperature cross-correlation functions in a
cylindrical cell \cite{shang04pre}.

\section{Summary and conclusion}
\label{sum}

To summarize, we have carried out a comparative study on local
mixing of active and passive scalars in turbulent thermal
convection. The statistical properties of local temperature are
experimentally compared with those of local concentration. It is
found that, although transported by the same turbulent flow field,
the fluctuations of the two scalars are distributed differently: a
decreasing exponential tail is found for the fluctuating
concentration, while Gaussian-like tails for the fluctuating
temperature. It is also shown that the temperature increments
exhibit a positive and large correlation with the concentration
increments above the Bolgiano time scale $t_B$; whereas this
correlation decays with decreasing time lags $\tau$ for
$\tau\lesssim t_B$. This reflects the fact that there is a certain
buoyant scale in turbulent RBC, above which the buoyancy effects
are important. Above $t_B$, temperature is active and the mixing
of both scalars are isotropic even though buoyancy acts on the
fluid in the vertical direction. While temperature fluctuations
are found to be more intermittent than concentration fluctuations,
suggesting that the active scalar possesses a higher level of
intermittency in turbulent thermal convection. Below $t_B$,
temperature is passive but its mixing are more anisotropic than
those of concentration, reflecting the fact that its diffusivity
($\kappa_T$) is larger than that of concentration ($\kappa_C$). It
is further found, from the simultaneous measurements of
temperature and concentration, that the two scalars have similar
autocorrelation functions and there is a strong and positive
correlation between them. Future investigations will be focused on
statistical and geometric features of the passive scalar field for
varying Ra.

\begin{acknowledgments}
We thank S.-Q. Zhou for making the temperature data in a
$81\times20\times81$-cm$^3$ rectangular cell available to us and
are grateful to E.S.C. Ching, D. Lohse and P.-E. Roche for many
valuable discussions. This work has been supported by the Research
Grants Council of Hong Kong SAR under Grant No. CUHK 403003 and
403806.
\end{acknowledgments}


\begin{thebibliography}{}

\bibitem[Shraiman and Siggia (2000)]{ss00nature}
\textsc{B.-I. Shraiman, E.-D. Siggia}, \textit{Nature}
\textbf{405}, 639, (2000).


\bibitem[Falkovich et al. (2001)]{falkovich01rmp}
\textsc{G. Falkovich, K. Gawedzki, M. Vergassola}, \textit{Rev.
Mod. Phys.} \textbf{73}, 913, (2001).

\bibitem[Castaing et-al. (1989)]{castaing89jfm}
\textsc{B. Castaing, G. Gunaratne, F. Heslot, L. Kadanoff, A.
Libchaber, S. Thomae, X.-Z. Wu, S. Zaleski, G. Zanetti},
\textit{J. Fluid Mech.} \textbf{204}, 1, (1989).

\bibitem[Sigga (1994)]{sigga94arfm}
\textsc{E.-D. Sigga}, \textit{Annu. Rev. Fluid Mech.} \textbf{26},
137, (1994).

\bibitem[Niemela et-al. (2000)]{sreenivasan00nature}
\textsc{J.-J. Niemela, L. Skrbek, K.-R. Sreenivasan, R.-J.
Donnelly}, \textit{Nature} \textbf{404}, 837, (2000).

\bibitem[Grossmann and Lohse (2000)]{gl00jfm}
\textsc{S. Grossmann, D. Lohse}, \textit{J. Fluid Mech.}
\textbf{407}, 27, (2000).

\bibitem[Kadanoff (2001)]{kadanoff01pt}
\textsc{L.-P. Kadanoff}, \textit{Phys. Today} \textbf{54}, 34,
(2001).

\bibitem[Sreenivasan (1991)]{sreenivasan91prsla}
\textsc{K.-R. Sreenivasan}, \textit{Proc. R. Soc. Lond. A}
\textbf{434}, 165, (1991).

\bibitem[Warhaft (2000)]{warhaft00arfm}
\textsc{Z. Warhaft}, \textit{Annv. Rev. Fluid Mech.} \textbf{32},
203, (2000).

\bibitem[Celani et-al. (2004)]{celani04njp}
\textsc{A. Celani, M. Cencini, A. Mazzino, M. Vergassola},
\textit{New J. Phys.} \textbf{6}, 72, (2004).

\bibitem[Celani et-al. (2002a)]{celani02prla}
\textsc{A. Celani, T. Matsumoto, A. Mazzino, M. Vergassola},
\textit{Phys. Rev. Lett.} \textbf{88}, 054503, (2002).

\bibitem[Celani et-al. (2002b)]{celani02prlb}
\textsc{A. Celani, M. Cencini, A. Mazzino, M. Vergassola},
\textit{Phys. Rev. Lett.} \textbf{89}, 234502, (2002).

\bibitem[Ching et-al. (2003)]{ching03pre}
\textsc{E.-S.-C. Ching, Y. Cohen, T. Gilbert, I. Procaccia},
\textit{Phys. Rev. E} \textbf{67}, 016304, (2003).

\bibitem[Gilbert and Mitra (2004)]{gilbert04pre}
\textsc{T. Gilbert, D. Mitra}, \textit{Phys. Rev. E} \textbf{69},
057301, (2004).

\bibitem[Zhou and Xia (2002)]{zhou02prl}
\textsc{S.-Q. Zhou, K.-Q. Xia}, \textit{Phys. Rev. Lett.}
\textbf{89}, 184502, (2002).

\bibitem[Xia et-al. (2003)]{xia03pre}
\textsc{K.-Q. Xia, C. Sun, S.-Q. Zhou},  \textit{Phys. Rev. E}
\textbf{68}, 066303, (2003).

\bibitem[Zhou et-al. (2007)]{zhou07pre}
\textsc{S.-Q. Zhou, C. Sun, K.-Q. Xia}, \textit{Phys. Rev. E}
\textbf{76}, 036301, (2007).

\bibitem[Sun et-al. (2008)]{sun07jfm}
\textsc{C. Sun, Y.-H. Cheung, K.-Q. Xia}, \textit{J. Fluid Mech.}
\textbf{submitted} (2008).

\bibitem[Koochesfahani and Dimotakis (1985)]{dimotakis85aiaa}
\textsc{M.-M. Koochesfahani, P.-E. Dimotakis}, \textit{AIAA J.}
\textbf{23}, 1700, (1985).

\bibitem[Walker (1987)]{walker87jpe}
\textsc{D.-A. Walker},  \textit{J. Phys. E: Sci. Instr.}
\textbf{20}, 217, (1987).

\bibitem[Prasad and Sreenivasan (1990)]{sreenivasan90jfm}
\textsc{R.-R. Prasad, K.-R. Sreenivasan}, \textit{J. Fluid Mech.}
\textbf{216}, 1, (1990).

\bibitem[Saylor (1995)]{saylor95eif}
\textsc{J.-R. Saylor}, \textit{Exp. Fluids} \textbf{18}, 445,
(1995).

\bibitem[Catrakis and Dimotakis (1996)]{dimotakis96jfm}
\textsc{H.-J. Catrakis, P.-E. Dimotakis}, \textit{J. Fluid Mech.}
\textbf{317}, 369, (1996).

\bibitem[Crimaldi (1997)]{crimaldi97eif}
\textsc{J.-P. Crimaldi}, \textit{Exp. Fluids} \textbf{23}, 325,
(1997).

\bibitem[Wang and Fiedler (2000)]{wang00eif}
\textsc{G.-R. Wang, H.-E. Fiedler}, \textit{Exp. Fluids}
\textbf{29}, 257, (2000).

\bibitem[Funfschilling et-al. (2005)]{ahlers05jfm}
\textsc{D. Funfschilling, E. Brown, A. Nikolaenko, G. Ahlers},
\textit{J. Fluid Mech.} \textbf{536}, 145, (2005).

\bibitem[Sun et-al. (2005a)]{sun05jfm}
\textsc{C. Sun, L.-Y. Ren, H. Song, K.-Q. Xia}, \textit{J. Fluid
Mech.} \textbf{542}, 165, (2005).

\bibitem[Ahlers et-al. (2006)]{ahlers06jfm}
\textsc{G. Ahlers, E. Brown, F.-F. Araujo, D. Funfschilling, S.
Grossmann, D. Lohse}, \textit{J. Fluid Mech.} \textbf{569}, 409,
(2006).

\bibitem[Sun et-al. (2005b)]{sun05prl}
\textsc{C. Sun, H.-D. Xi, K.-Q. Xia}, \textit{Phys. Rev. Lett.}
\textbf{95}, 074502, (2005).

\bibitem[Brown et-al. (2005)]{ahlers05prl}
\textsc{E. Brown, A. Nikolaenko, G. Ahlers}, \textit{Phys. Rev.
Lett.} \textbf{95}, 084503, (2005).

\bibitem[Xi et-al. (2006)]{xi06pre}
\textsc{H.-D. Xi, Q. Zhou, K.-Q. Xia}, \textit{Phys. Rev. E}
\textbf{73}, 056312, (2006).

\bibitem[Qiu and Xia (1998)]{qiu98pre}
\textsc{X.-L. Qiu, K.-Q. Xia}, \textit{phys. Rev. E} \textbf{58},
486, (1998).

\bibitem[Jayesh and Warhaft (1991)]{warhaft91prl}
\textsc{Jayesh, Z. Warhaft}, \textit{Phys. Rev. Lett.}
\textbf{67}, 3503, (1991).

\bibitem[Qiu et-al. (2004)]{qiu04pof}
\textsc{X.-L. Qiu, X.-D. Shang, P. Tong, K.-Q. Xia}, \textit{Phys.
Fluids} \textbf{16}, 412, (2004).

\bibitem[Sun and Xia (2005)]{sun05pre}
\textsc{C. Sun, K.-Q. Xia}, \textit{Phys. Rev. E} \textbf{72},
067302, (2005).

\bibitem[Tavoularis and Corrsin (1981)]{tavoularis81jfm}
\textsc{S. Tavoularis, S. Corrsin}, \textit{J. Fluid Mech.}
\textbf{104}, 311, (1981).

\bibitem[Villermaux et-al. (1998)]{villermaux98crass}
\textsc{E. Villermaux, C. Innocenti, J. Duplat}, \textit{C. R.
Acad. Sci. Paris} \textbf{326}, 21, (1998).

\bibitem[Solomon and Gollub (1991)]{gollub91pra}
\textsc{T.-H. Solomon, J.-P. Gollub}, \textit{Phys. Rev. A}
\textbf{43}, 6683, (1991).

\bibitem[Kolmogorov (1941a)]{kolmogorov1941a}
\textsc{A.-N. Kolmogorov}, \textit{Dokl. Akad. Nauk SSSR}
\textbf{30}, 229, (1941).

\bibitem[Kolmogorov (1941b)]{kolmogorov1941b}
\textsc{A.-N. Kolmogorov}, \textit{Dokl. Akad. Nauk SSSR}
\textbf{32}, 19, (1941).

\bibitem[Obukhov (1949)]{O49}
\textsc{A.-M. Obukhov}, \textit{Izv. Akad. Nauk. SSSR Geogr.
Geofiz.} \textbf{13}, 58, (1949).

\bibitem[Corrsin (1951)]{C51}
\textsc{S. Corrsin}, \textit{J. Appl. Phys.} \textbf{22}, 469,
(1951).

\bibitem[R. Bolgiano (1959)]{B59}
\textsc{R. Bolgiano}, \textit{J. Geophys. Res.} \textbf{64}, 2226,
(1959).

\bibitem[A.M. Obukhov (1959)]{O59}
\textsc{A.M. Obukhov}, \textit{Dokl. Akad. Nauk. SSSR}
\textbf{125}, 1246, (1959).

\bibitem[Sun et-al. (2006)]{sun06prl}
\textsc{C. Sun, Q. Zhou, K.-Q. Xia}, \textit{Phys. Rev. Lett.}
\textbf{97}, 144504, (2006).

\bibitem[Shang and Xia (2001)]{shang01pre}
\textsc{X.-D. Shang, K.-Q. Xia}, \textit{Phys. Rev. E}
\textbf{64}, 065301(R), (2001).

\bibitem[Wu et-al. (1990)]{wu90prl}
\textsc{X.-Z. Wu, L. Kadanoff, A. Libchaber, M. Sano},
\textit{Phys. Rev. Lett.} \textbf{64}, 2140, (1990).

\bibitem[Du and Tong (2001)]{tong01pre}
\textsc{Y.-B. Du, P. Tong}, \textit{Phys. Rev. E} \textbf{63},
046303, (2001).

\bibitem[Monin and Yaglom (1975)]{my75}
\textsc{A.-S. Monin, A.-M. Yaglom}, \textit{Statistical fluid
mechanics.} MIT Press, (1975).

\bibitem[Ching et-al. (2004)]{ching04jot}
\textsc{E.-S.-C. Ching, K.-W. Chui, X.-D. Shang, X.-L. Qiu, P.
Tong, K.-Q. Xia}, \textit{J. Turbulence} \textbf{5}, 027, (2004).

\bibitem[Gasteuil et-al. (2007)]{pinton07prl}
\textsc{Y. Gasteuil, W.-L. Shew, M. Gibert, B.-C. F.-Chill\'{a},
J.-F. Pinton}, \textit{Phys. Rev. Lett.} \textbf{99}, 234302,
(2007).

\bibitem[Schumacher and Sreenivasan (2003)]{sreenivasan03prl}
\textsc{J. Schumacher, K.-R. Sreenivasan}, \textit{Phys. Rev.
Lett.} \textbf{91}, 174501, (2003).

\bibitem[Benzi et-al. (1993a)]{benzi93a}
\textsc{R. Benzi, S. Ciliberto, R. Tripiccione, C. Bauder, F.
Massaioli, S. Succi}, \textit{Phys. Rev. E} \textbf{48}, 29(R),
(1993).

\bibitem[Benzi et-al. (1993b)]{benzi93b}
\textsc{R. Benzi, S. Ciliberto, C. Bauder, G. Ruiz-Chavarria, R.
Tripiccione}, \textit{Eurphys. Lett.} \textbf{24}, 275, (1993).

\bibitem[Moisy et-al. (2001)]{tabeling01prl}
\textsc{F. Moisy, H. Willaime, J.-S. Andersen, P. Tabeling},
\textit{Phys. Rev. Lett.} \textbf{86}, 4827, (2001).

\bibitem[Shang et-al. (2004)]{shang04pre}
\textsc{X.-D. Shang, X.-L. Qiu, P. Tong, K.-Q. Xia}, \textit{Phys.
Rev. E} \textbf{70}, 026308, (2004).


\end{thebibliography}
\end{document}